\documentclass[article]{aa}

\usepackage[varg]{txfonts}
\usepackage{mathtools}
\usepackage{url}
\usepackage[breaklinks=true]{hyperref}
\usepackage{twoopt}
\usepackage{natbib}
\bibpunct{(}{)}{;}{a}{}{,} 
\usepackage{ctable}
\usepackage{multirow}
\usepackage{xcolor}  
\usepackage{enumitem}
\usepackage{xcolor}
\usepackage{footmisc}
\hypersetup{colorlinks=true,citecolor=blue} 

\newcommand{\thour}[1]{{#1}$^{\mathrm{h}}$}
\newcommand{\tmin}[1]{{#1}$^{\mathrm{m}}$}
\newcommand{\tsec}[1]{{#1}$^{\mathrm{s}}$}
\newcommand{\thms}[3]{\thour{#1}\tmin{#2}\tsec{#3}}
\newcommand{\adeg}[1]{{#1}$^{\circ}$}
\newcommand{\amin}[1]{{#1}$^\prime$}
\newcommand{\asec}[1]{{#1}$^{\prime\prime}$}
\newcommand{\adms}[3]{\adeg{#1}\amin{#2}\asec{#3}}

\newcommand{\sbeam}[2]{\asec{#1}$\times\,$\asec{#2}}

\newcommand{\vla}{\emph{VLA}}
\newcommand{\jvla}{\emph{JVLA}}

\newcommand{\gmrt}{GMRT}
\newcommand{\binary}{Ross~867-8}
\newcommand{\lofar}{LOFAR}

\newcommand{\gaia}{\textit{Gaia}}

\newcommand{\stokesi}{\textit{I}}

\newcommand{\stokesv}{\textit{V}}
\newcommand{\stokesrr}{\textit{RR}}
\newcommand{\stokesll}{\textit{LL}}

\newcommand{\stokeschalf}{\textit{RR,LL}}
\newcommand{\stokescfull}{\textit{RR,LL,RL,LR}}

\newcommand{\stokeslfull}{\textit{XX,YY,XY,YX}}

\newcommand{\cluster}{RXJ1720.1+2638}
\newcommand{\target}{Ross~867}
\newcommand{\companion}{Ross~868}
\newcommand{\neighbor}{NVSS~J171949+263007}
\newcommand{\RNum}[1]{\uppercase\expandafter{\romannumeral #1\relax}}

\begin{document}

\title{Differences in radio emission from similar M dwarfs \\ in the binary system~\binary{}}
\titlerunning{Radio emission differences from identical M dwarfs~\binary{}}

\author{L.~H.~Quiroga-Nu\~{n}ez\inst{1,2}
\and H.~T.~Intema\inst{1,3}
\and J.~R.~Callingham\inst{4}
\and J.~Villadsen\inst{5}
\and H.~J.~van~Langevelde\inst{2,1}
\and \\ P.~Jagannathan\inst{6}
\and T.~W.~Shimwell\inst{4,1}
\and E.~P.~Boven\inst{2,1}
} 
\authorrunning{L.H. Quiroga-Nu\~{n}ez et al.} 

\institute{Leiden Observatory, Leiden University, PO Box 9513, 2300 RA, Leiden, The Netherlands \\ 
\email{quiroganunez@strw.leidenuniv.nl}
\and Joint Institute for VLBI ERIC (JIVE), Oude Hoogeveensedijk 4, 7991 PD Dwingeloo, The Netherlands
\and International Centre for Radio Astronomy Research -- Curtin University, GPO Box U1987, Perth, WA 6845, Australia
\and ASTRON, Netherlands Institute for Radio Astronomy, Oude Hoogeveensedijk 4, 7991 PD, Dwingeloo, The Netherlands
\and National Radio Astronomy Observatory, 520 Edgemont Rd., Charlottesville, VA 22903, USA
\and National Radio Astronomy Observatory, 1003 Lopezville Road, Socorro, NM 87801-0387, USA}
\date{Received ..., 2019; accepted ..., 2019}

\abstract{Serendipitously, we rediscovered radio emission from the binary system~\target{} (M4.5V) and~\companion{} (M3.5V) while inspecting archival Giant Metrewave Radio Telescope (GMRT) observations. The binary system consists of two M-dwarf stars that share common characteristics such as spectral type, astrometric parameters, age, and emission at infrared, optical, and X-ray frequencies. The~\gmrt{} data at 610 MHz taken on July 2011 shows that the radio emission from~\target{} is polarized and highly variable on hour timescales with a peak flux of 10.4 $\pm$ 0.7 mJy/beam. Additionally, after reviewing archival data from several observatories (\vla{}, \gmrt{}, \jvla{}, and~\lofar{}), we confirm that although the two stars are likely coeval, only~\target{} was detected, while~\companion{} remains undetected at radio wavelengths. As the stars have a large orbital separation, this binary stellar system provides a coeval laboratory to examine and constrain the stellar properties linked to radio activity in M dwarfs. We speculate that the observed difference in radio activity between the dwarfs could be due to vastly different magnetic field topologies or that~\target{} has an intrinsically different dynamo.}

\maketitle 


\section{Introduction}
\label{sec:intro}

The most common stellar type is an M dwarf~\citep{Henry2006}, and understanding eruptive events in these stars (e.g., stellar flares and stellar coronal mass ejections) is a fundamental astrophysical subject, in particular, for assessing the habitability of orbiting exoplanets~\citep{Crosley2018} that have already been detected~\citep{Dressing2013}. It is known that these stars show dynamic activity in their stellar atmospheres, especially for mid- to late-M dwarfs~\citep[e.g.,][]{West2008a}. 
This effect has usually been attributed to magnetic energy release in the outer atmosphere that accelerates particles and gives rise to chromospheric and coronal heating, producing photons at all wavelengths~\citep{Bochanski2007}. In particular, at radio wavelengths the flux intensity changes are not well understood, and although the physical mechanisms that produce this radio variability have been studied in the Sun~\citep[e.g.,][]{Shibata2011a}, recent studies show that radio flares coming from M dwarfs do not have a clear solar analog~\cite[see, e.g.,][and the references within]{Villadsen2019}.
 In order to establish the physical mechanisms occurring in M dwarfs that could lead to radio intensity changes, previous investigations have analyzed a wide range of stellar properties~\citep[][]{White1989,Berger2002,Wright2011,Houdebine2015,Houdebine2017,Newton2017,Yang2017a}. Factors such as the stellar spectral type, the Rossby number, 
 and ages of dwarf stars may be related to the magnetic activity and therefore dynamic radio activity~\citep[][]{West2008a,Lopez-Santiago2010,Mclean2012,West2015,Barnes2017,Ilin2019}.
 
 \begin{table}[]
\caption{Astrophysical information of the binary system~\binary{}.\label{t_sourcesinfo}}
\begin{center}
\resizebox{\hsize}{!}{ 
\begin{tabular}{lrr}
\hline
\hline
                  & Ross 867            & Ross 868            \\
\hline
Spectral Type\tmark[(1)]  & M4.5\RNum{5}   & M3.5\RNum{5}   \\
Mass\tmark[(2)]  ($\rm{M_{\odot}}$)& 0.311 $\pm$ 0.132 & 0.376 $\pm$ 0.040\\
Parallax\tmark[(3)] (mas)                               & $92.967 \pm 0.061$                & $92.988 \pm 0.050$                \\
Distance\tmark[(3,14)] (pc)                               & $10.753 \pm 0.007$                & $10.751 \pm 0.006$                \\
Radius ($\rm{R_{\odot}}$)& [0.27-0.51]\tmark[(9,7)] & [0.478-0.535]\tmark[(6,4)]\\
$\rm{V_{LSR}\tmark[(1)] \ (km \ s^{-1})}$      & $-34.6 \pm 0.2$                 & $-34.9 \pm 0.1$                 \\
$\mu_{\alpha}$\tmark[(3)] $\rm{(mas \ s^{-1})}$         & $-226.1 \pm 0.1$                & $-214.8 \pm 0.1$                \\
$\mu_{\delta}$\tmark[(3)] $\rm{(mas \ s^{-1})}$         & $355.3\pm 0.1$                  & $351.0\pm 0.1$                  \\
Age\tmark[(12)] (Myr)                                    & [90-300]                    & [25-300]                    \\
$\rm{T_{eff}}$ (K)                           & 2667\tmark[(2)]                         & 3319 $\pm$ 100\tmark[(6)]                            \\
$ V sin \ i$ $\rm{(km \ s^{-1})}$      & [6.79-10]\tmark[(5,8)]                            & [1.0-28.5]\tmark[(4,6,7,10,11,12,13)]                           \\
B\tmark[(1)] (mag)          & 14.67                           & 12.98                           \\
R\tmark[(1)] (mag)          & $12.65\pm 0.06$                 & $11.34\pm 0.02$                 \\
G\tmark[(1)] (mag)          & $11.456\pm 0.001$               & $10.137\pm 0.001$               \\
J\tmark[(1)] (mag)          & $8.23\pm 0.02$                  & $7.27\pm 0.02$                  \\
H\tmark[(1)] (mag)          & $7.64\pm 0.03$                  & $6.71\pm 0.03$                  \\
K\tmark[(1)] (mag)          & $7.35\pm 0.03$                  & $6.42\pm 0.02$                  \\
XMM\tmark[(13)] {[}0.2-12keV{]}                         & \multirow{2}{*}{$3.4\pm 0.1$}   & \multirow{2}{*}{$2.1\pm 0.1$}   \\
$\rm{(10^{-12} \ erg \ s^{-1} \ cm^{-2})}$ &                                 &                                 \\
Chandra\tmark[(13)]  {[}0.3-11keV{]}                     & \multirow{2}{*}{$1.26\pm 0.03$} & \multirow{2}{*}{$1.25\pm 0.02$} \\
$\rm{(10^{-12} \ erg \ s^{-1} \ cm^{-2})}$ &                                 &                                \\
\hline
\end{tabular}
}
\end{center}
{\footnotesize {
\bf Notes.} When several measurements differ, a range in square brackets is given. {\bf References.} (1) The SIMBAD astronomical database~\citep{Wenger2000}. (2) \cite{Jenkins2009}. (3) \cite{Brown2018}. (4) \cite{Houdebine2012}. (5) \cite{Jeffers2018}. (6) \cite{Houdebine2016}. (7) \cite{Caillault1990}. (8) \cite{Mclean2012}. (9) \cite{White1989}. (10) \cite{Moutou2017}. (11) \cite{Reiners2012}. (12) \cite{Shkolnik2012}. (12) \cite{Kiraga2007}. (13) This work. (14) Assuming geometrical distance calculation from parallax made by~\cite{Bailer-Jones2018a}.}
\end{table}

In this paper we present a serendipitous flare detection of the variable radio source~\target{} (also known as V639~Her or Gliese~669B) based on archival observations from the Giant Metrewave Radio Telescope~\citep[\gmrt{};][]{Swarup1991} that targeted the galaxy cluster~\cluster{}~\citep{Giacintucci2014,Savini2019}. 
Although~\target{} had already been marked as radio loud by~\cite{Jackson1987}, there was no evidence of variability or circular polarized emission reported. \target{} is part of a binary system with~\companion{} (also known as V647~Her, Gliese~669A or HIP 84794); they are separated by 179.3 $\pm$ 0.1 AU (or 16.854 $\pm$ 0.001\asec{}) in projection, and both stars have been classified as optical flaring stars~\citep{Samus2004,Nakajima2010a}. Using their stellar position together with recent parallax measurements in {\it Gaia} DR2~\citep{Brown2018}, we confirm that the two sources are close enough to be gravitationally bound, but separate enough to discard any stellar material transfer or significant tidal interaction. What is remarkable about this binary system is that although the stars are similar in terms of spectral type, age, and high 3D motion, as well as IR, optical (the same thermal emission processes can be assumed), and X-ray emission (see Table~\ref{t_sourcesinfo}), only~\target{} seems to be radio loud. Hence, this binary stellar system provides a coeval laboratory to examine and constrain the stellar properties present in radio-loud flare stars, and hence to discuss the physical mechanisms behind the radio flare emission in M dwarfs. 
In order to study this binary system in more detail, we initiated a search for other historical radio data identifying several radio continuum observations in the archives of the~\gmrt{}, the Very Large Array~\citep[\vla{};][]{Thompson1980}, and the Low Frequency Array~\citep[\lofar{};][]{VanHaarlem2013} observed at different epochs and different radio frequencies. In this paper we present an analysis of this set of observations, confirming that~\target{} is indeed a source of time-variable, bright radio emission, whereas~\companion{} remains undetected. 
\section{Observations}\label{sec:observations}

The~\gmrt{} observed the galaxy cluster~\cluster{} four times at various frequencies, one of which is a dual-frequency observation (project codes 11MOA01 and 20\_016; see Tables~\ref{tab:gmrt_measurements} and~\ref{tab:observations} for details). In three observations the star~\target{} was located within the field of view. 
\gmrt{} observations amounted to integrations of four to six hours, consisting predominantly of blocks of $\sim$30 minutes on~\cluster{} interleaved with phase calibrator observations of several minutes (see details in Table~\ref{tab:gmrt_measurements}). All but the dual-frequency observation recorded instrumental~\stokesrr{} and~\stokesll{} visibilities, while the dual-frequency observations recorded~\stokesrr{} visibilities only at 610~MHz, and~\stokesll{} visibilities only at 235~MHz. Full polarization calibration was not possible for any of these data sets. 
However, under the general assumptions that the circular feeds are mostly orthogonal with around $\lesssim 10\%$ leakage per antenna~\citep{Joshi2010,Roy2010} and that the sources are not linearly polarized at these low frequencies~\citep[Faraday rotation would wipe out the linearly polarized light given the density of the plasma,][]{Dulk1985,White1995}, \stokesrr{} and~\stokesll{} visibilities can be converted into Stokes~\stokesi{} and~\stokesv{} visibilities. For the dual-frequency observations, the individual~\stokesrr{} and~\stokesll{} visibilities are approximately equal to Stokes~\stokesi{} visibilities, assuming that the radio sky emission is predominantly circularly unpolarized (which is generally the case except for gyro-synchrotron emitters such as planets, stars, and the Galactic center) and that the instrumental polarization over the relevant part of the field of view is negligible~\citep{Wielebinski2012,Farnes2014}. 

\begin{table*}[tbp]
\caption{Measured radio properties of~\target{} using \gmrt{} data.\label{tab:gmrt_measurements}}
\begin{center}
\resizebox{\hsize}{!}{ 
\begin{tabular}{lcccccccc}
\hline \hline
Date (yyyy-mm-dd) & Frequency & Blocks $\times$ time& Observational & Correlation & $\alpha \pm \Delta\alpha$ & $ \delta \pm \Delta\delta$ \tmark[(1)] & $S_{\nu}$\tmark[(2)] & Ref $S_{\nu}$\tmark[(3)] \\
and UTC range & (MHz) & per block (min) & time (min) & parameters & (arcsec) & (arcsec) & (mJy) & (mJy) \\
\hline
2007-03-08 & 325 & 7.5 $\times$ 37 & 275 & \stokesi & $52.87 \pm 0.11$ & $ 05.7\pm 0.11$ & $5.3 \pm 0.8$ & $57.0 \pm 5.7$ \\
23:28-06:11(+1) & & & & \stokesll & $52.84\pm 0.05$ & $ 05.6 \pm 0.07$ & $11.6 \pm 1.2$ & $67.2 \pm 6.7$ \\
& & & & \stokesrr & $52.84\pm0.19$ & $ 05.7\pm0.21$ & $3.5 \pm 0.6$ & $62.2 \pm 6.2$ \\
2007-03-10 & 610 & 12 $\times$ 26 & 313 & \stokesi & $52.85 \pm 0.08$ & $ 05.4 \pm 0.10$ & $1.7 \pm 0.2$ & $39.1 \pm 3.9$ \\
23:17-06:09(+1)& & & & \stokesll & $52.83 \pm 0.09$ & $ 05.1\pm 0.08$ & $2.4 \pm 0.3$ & $41.0 \pm 4.1$ \\
 & & & & \stokesrr & $52.80\pm0.19$ & $ 05.7\pm 0.26$ & $0.9 \pm 0.2$ & $38.0 \pm 3.8$ \\
2011-07-24 & 235 & 11 $\times$ 31 & 346 & \stokesll & $52.78 \pm 0.23$ & $ 06.6\pm 0.21$ & $6.8 \pm 0.6$ & $96.2 \pm 9.7$ \\
13:30-21:25& 610 & & & \stokesrr & $52.76\pm 0.01$ & $ 06.9 \pm 0.01$ & $7.8 \pm 0.8$ & $39.6 \pm 4.0$ \\
\hline
\end{tabular}}
\end{center}
{\footnotesize {\bf Notes.} (1) Source position and uncertainty of right ascension ($\alpha$) and declination ($\delta$) measured from~\thms{17}{19}{00} and~\adms{+26}{30}{00} in sexagesimal notation. (2) Source flux density and uncertainty. (3) Reference flux density and uncertainty of neighboring source~\neighbor{}.
}
\end{table*}

We used the SPAM pipeline~\citep{Intema2017} to process the archival~\gmrt{} observations (see Appendix~\ref{SPAM} for a concise description) 
and generated calibrated Stokes~\stokesi{} visibilities and images for all observations. Additionally, to look for evidence of circular polarization in observations with both~\stokesrr{} and~\stokesll{} present, we split off each polarization and processed them independently using the same pipeline. The resulting sensitivities and resolutions of the wide-field images generated by the pipeline are listed in Table~\ref{tab:observations}. 

The source extraction software PyBDSF~\citep{Mohan2015} was used to measure the position and flux density of all detectable sources in the primary-beam-corrected images. The measured positions and flux densities of~\target{} are given in Table~\ref{tab:gmrt_measurements}. The flux density calibration procedure for~\gmrt{} is typically accurate to about 10~\%~\citep[e.g.,][]{Chandra2004}, so a systematic 10~\%\ error is added (in quadrature) to all the random flux uncertainties as reported by PyBDSF (and to all~\gmrt{} flux measurements in the remainder of the article). Table~\ref{tab:gmrt_measurements} also contains the flux densities of the bright neighboring extragalactic source~\neighbor{} (\amin{0.8} west of~\target{}), which we use as a reference source. The reported 1.4~GHz flux densities of this source in the NVSS and FIRST surveys are $17.3 \pm 0.6$ and $17.9 \pm 0.9$~mJy, respectively (see Table~\ref{t_first_nvss} for details). Given the difference in resolution and observing epochs, this indicates that the reference source is compact and not variable on short timescales (months to years). The position of the reference source is measured to be~\thms{17}{19}{49.31}~\adms{+26}{30}{07.7} in all \gmrt{} images within~\asec{0.1} accuracy.


\section{Results}
\label{sec:results}

\subsection{Change in flux and position of~\target{}}

By comparing two separate observations at 610~MHz from 2007 and 2011, we discovered that one of the radio sources shifted its position and changed its flux density between the epochs. After aligning the observations using the close-by quasar~\neighbor{} (at~\amin{0.8}) as astrometric reference source, we found that the variable radio emission unambiguously coincided with the sky position of~\target{} in both observations. The radio properties obtained for~\target{} and~\neighbor{} are described in Table~\ref{tab:gmrt_measurements}, which reveals several important properties.

First, there is a noticeable shift in the position of~\target{} between March 2007 and July 2011, which is much larger than the expected parallax shift expected (\asec{0.1}) for this stellar source. When using the 610~MHz positions (highest resolution) in the \stokesll{} images (the only polarization available at both epochs), the shift is~\asec{$-0.6$} in right ascension and~\asec{1.2} in declination over 4.37~yr, which translates to a proper motion of $\rm{-137 \pm \ 44 \ and \ 275 \ \pm \ 60 \ mas \ yr^{-1}}$ for right ascension and declination, respectively. These values differ by nearly two sigma with respect to the recently estimated values by { \it Gaia} DR2 (Table~\ref{t_sourcesinfo}). The discrepancy is likely due to the astrometric measurement uncertainties caused by the limited resolution and signal-to-noise ratio in \gmrt{}.
 
Second, there is a significant increase in the flux density of~\target{} at 610~MHz between March 2007 and July 2011. The flux ratio lies in the range of 3--10 taking into account the flux uncertainties and the Stokes measurement selected. The change in flux density over time of the quasar reference source (\neighbor{}) is negligible and lies within the uncertainties.

Third, there is a significant difference in the flux density of \target{} between the correlation parameters~\stokesrr{} and~\stokesll{} for the cases where both visibilities were recorded, namely at 325 and 610~MHz in March 2007. At both frequencies the flux density in~\stokesll{} is $3\pm1$ times higher than in~\stokesrr{} (see Table~\ref{tab:gmrt_measurements}). Whereas for the quasar reference source (\neighbor{}) the difference between the~\stokesrr{} and~\stokesll{} is consistent with unpolarized emission. This observational evidence demonstrates unambiguously that time-variable emission, likely circularly polarized radio emission, is originating from~\target{}.
\subsection{Radio light curves}
\label{sec:light_curve}

\begin{figure}
\resizebox{1.02\hsize}{!}{\includegraphics[width=8cm]{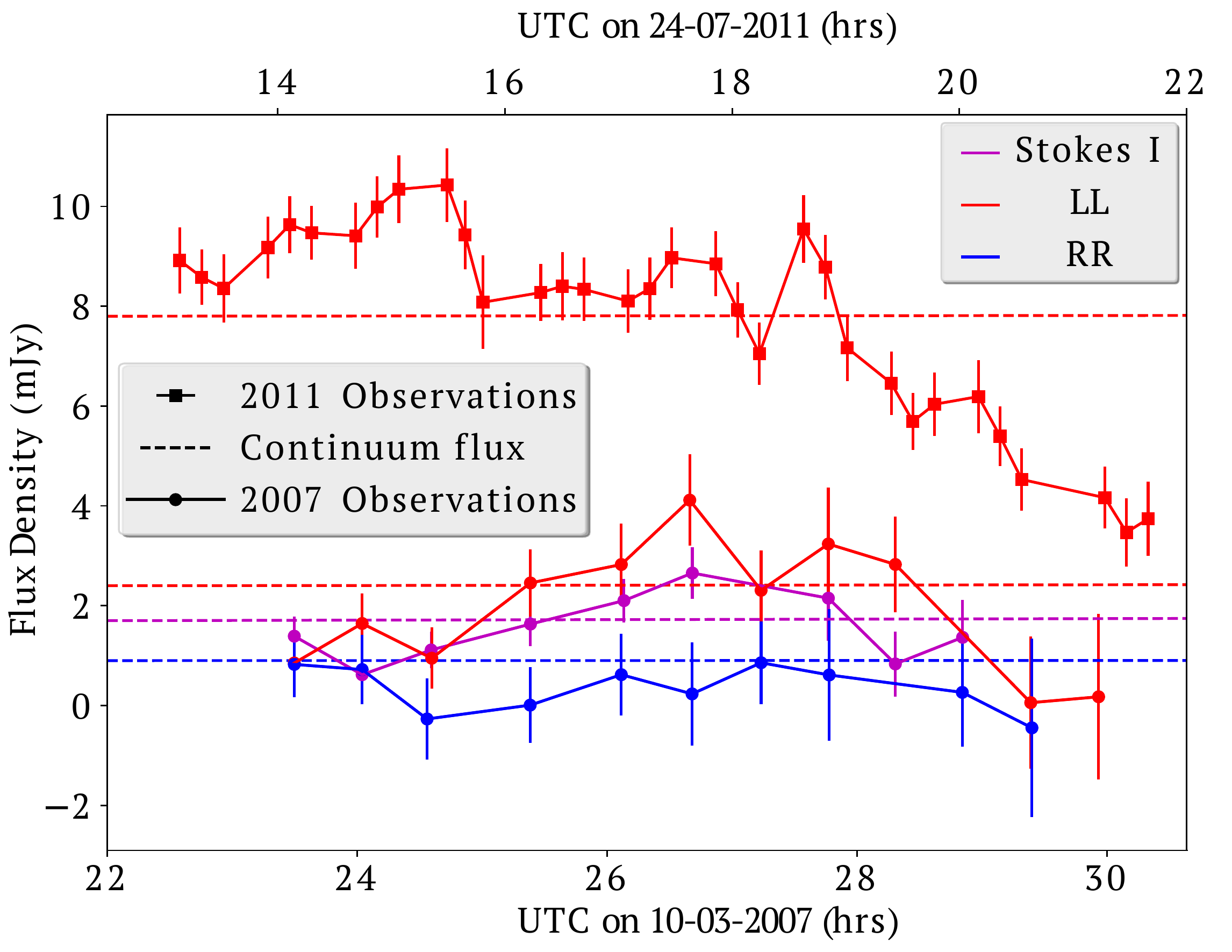}}
\caption{Comparison between radio light curves of \target{} as measured with the~\gmrt{} at 610~MHz during observations in 2007 (lower six curves and axis) and 2011 (upper two curves and axis). 
All neighboring points are connected with lines to help guide the eye. The vertical error bars represent the flux density measurement uncertainties. The time ranges of the snapshot image in which the flux measurement was made were $\sim$0.40 and $\sim$0.15 hours for the 2007 and 2011 observations, respectively, and are shown as horizontal error bars in the legend. The dashed horizontal lines indicate the average continuum flux density of~\target{} as measured in the image from the full time range.} 
\label{fig:light_curves_610}
\end{figure}

Since the~\gmrt{} observations are integrations of several hours, consisting of blocks of 0.5 hours, this allowed us to look for radio variability of~\target{} on timescales of minutes to hours, by reimaging time slices of the calibrated and flagged visibility data as generated by the pipeline. As we are only interested in~\target{} and the reference source~\neighbor{}, we pre-subtracted all radio source flux outside a~\amin{2} radius circle centered on~\target{} from the full visibility data set. 
When splitting an observation into $N$ time slices of equal duration, the theoretical sensitivity in a single time slice image is worse by a factor of $\sqrt{N}$ compared to the full time range. We used this scaling to determine the minimum size of each time slice and the depth to which we deconvolve (CLEAN) the time slice image. 

After making a time series of images per data set, we measured the flux density of~\target{} and~\neighbor{} by integrating over tight apertures (2.5 times the theoretical beam size of each image) around measured peak position for each source. 
By measuring the flux density in each time slice, we attempted to construct radio light curves of~\target{} for the available observations. This worked best for the~\gmrt{} observations at 610~MHz because of significant signal-to-noise ratio limitations in the other observations. At 610 MHz, the flux measurements on the reference source~\neighbor{} show a mild ($<$10 \%) variation over time, which can be attributed to limited image quality per snapshot and to higher-order primary-beam effects, as we explain here. First, our snapshots are very short observations yielding poor UV coverage (due to the Earth's rotation), which leads to less accurate flux density measurements. Second, instrumental amplitude variations of a few percentage points may be expected as the beam patterns of all antennas in the interferometer cannot be assumed to be the same, especially far from the the pointing center (\cluster{}) where Ross 867 and the reference source are located.

We corrected the snapshot flux density measurements of~\target{} by multiplying them with the full-to-snapshot flux ratios of the reference source; we make the assumption that its flux density does not change on a timescale of hours. Figure~\ref{fig:light_curves_610} shows the resulting light curves from the two available 610~MHz observations, with a peak flux of 10.4 $\pm$ 0.7 mJy/beam in 2011. The time resolution of the 2007 radio light curve is about three times worse than the 2011 radio light curve due to signal-to-noise ratio limitations. 

\section{Discussion}
\label{sec:discussion}

\begin{figure*}
\centering 
\resizebox{\hsize}{!}{\includegraphics[width=16cm]{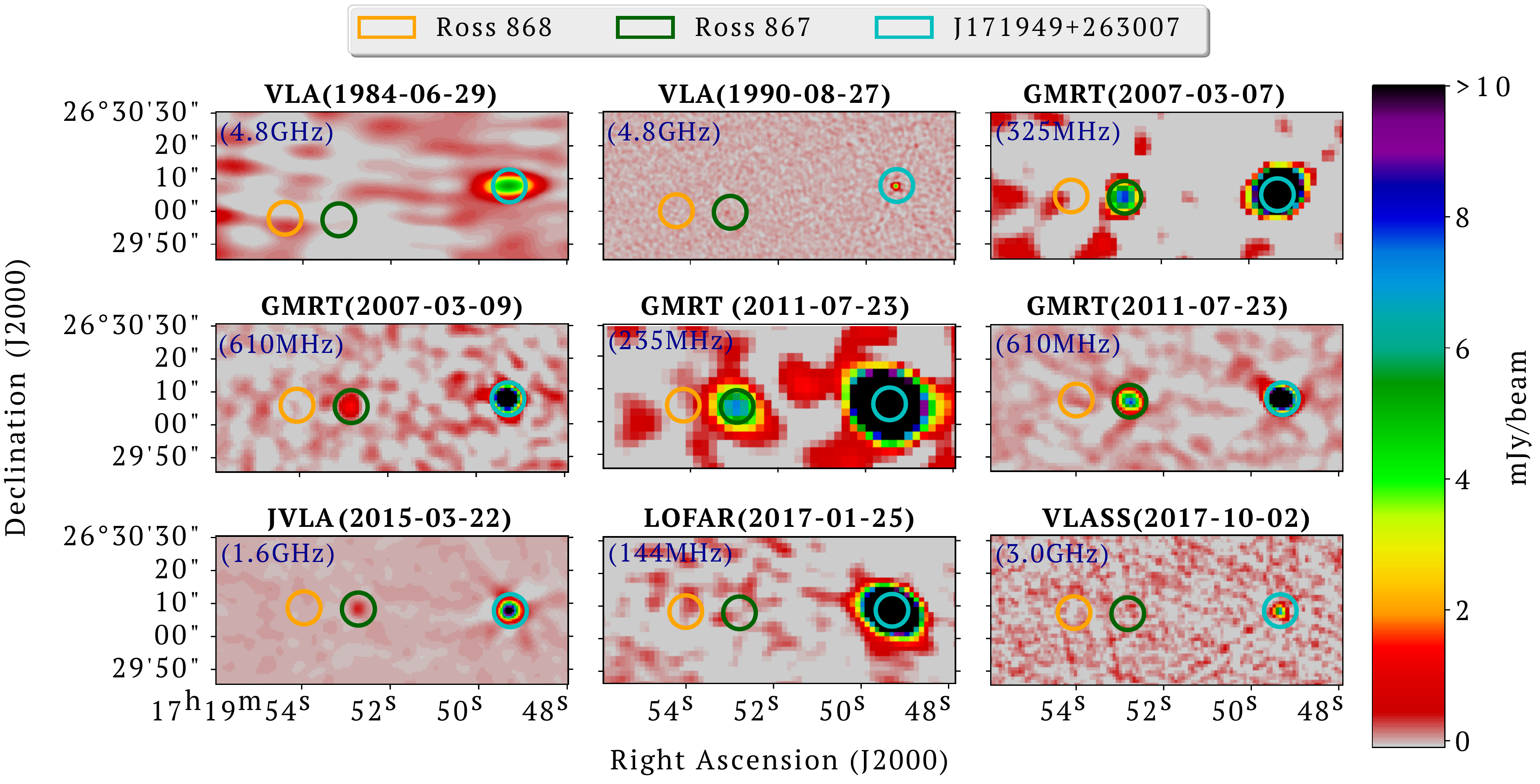}}
\caption{Multiple radio observations at different epochs and frequencies of the field of view that contains~\target{}, \companion{}, and~\neighbor{} (orange, green, and light blue apertures, respectively). Although~\target{} shows stellar activity at different epochs and frequencies, its similar binary (\companion{}) remains undetectable at all epochs and radio frequencies. The aperture positions were estimated assuming the proper motions measured by {\it Gaia} for each observed date. Table~\ref{tab:observations} gives the technical information for each data set inspected. \label{fig_vla_lofar}}
\end{figure*} 

\subsection{Identical stellar origin}

The recent~\gaia{} DR2 measurements (Table~\ref{t_sourcesinfo}) show that~\target{} and~\companion{} have equal distance within the errors. Moreover, their relative velocity is $\rm{0.7 \pm 0.3 \ km \ s^{-1}}$ and their separation is 179.3 $\pm$ 0.1 AU in projection. 
This large separation allows us to argue that they are not magnetically interacting since the magnetosphere of such stars cannot extend further than approximately five stellar radii~\citep{Gudel2002}. Therefore, their flare emissions can arguably be studied as independent events from similar stellar sources.
Furthermore, although the age estimates for~\binary{} have large margins (Table~\ref{t_sourcesinfo}), their similar peculiar 3D motion, spectral type, and orbital separation strongly suggest that both stars are coeval (likely related within the Hyades group;~\citealt{Nakajima2010a,Shkolnik2012}), which establishes another constraint for the flaring-age relation of dwarfs in open clusters~\citep{Ilin2019}.

\subsection{Nature of the flare emission}\label{nature}

Since the~\gmrt{} observations obtained in 2011 were only recorded at~\stokesll{} polarization, we cannot establish if the emission from~\target{} was circularly polarized at that epoch. However, the observations made in 2007 at 325 and 610 MHz with the~\gmrt{} indicate that the emission is indeed polarized (see Figure~\ref{fig:light_curves_610}). 
While we lack the cross-polarization terms in our 610 MHz observation, we can argue that stellar emission is likely not linearly polarized \citep[e.g.,][]{Villadsen2019}. With this assumption, we infer that the 610 MHz emission from~\target{} in 2011 is $\sim$40$\%$ circularly polarized. We also estimate that the brightness temperature of the emission is $\rm{>0.7\times10^{12} \ K}$ (assuming that the emission site is the entire photosphere) meaning that it is unlikely to be gyrosynchrotron emission~\citep{Melrose1982,Dulk1985}. Therefore, the emission coming from~\target{} is likely caused by a coherent process, either electron cyclotron maser instability or plasma emission. Furthermore, the emission detected in 2011 at 610 MHz shows that the flare emission lasted for at least seven hours, which is longer than usual for flares of this type~\citep{Lynch2017}, but still consistent with timescales for coherent emission from M dwarfs~\citep{Slee2003,Villadsen2019}. 
It is possible that~\target{} is already a fully convective star, explaining the polarized flare emission detected, whereas~\companion{} would still have a radiative core. However, we note that non-fully convective stars (e.g., the Sun) can also produce strongly polarized emission from plasma emission~\citep{Bastian1998}. 

\subsection{Historical radio observations}\label{sec:historical}

We have extensively reviewed archival data from several radio observatories for the field of view that contains~\binary{}. Table~\ref{tab:observations} displays technical information for each data set inspected, and Figure~\ref{fig_vla_lofar} shows the field of view from different radio observatories at several epochs and frequencies, where the expected positions at each observational time for~\target{} and~\companion{} (assuming the proper motion measured by~\gaia{} DR2) are highlighted. Since ~\companion{} was undetected in all observations, here we only discuss the~\target{} detection.

A candidate radio detection of~\target{} with Arecibo at 430~MHz ($\sim$\amin{10} resolution) was first reported by~\cite{Spangler1976a}, but given the high flux density reported ($320\pm 90$ mJy) and the vicinity of a bright extragalactic radio source (\neighbor{}) at a distance of~\amin{$0.8$} from \target{} as seen in higher resolution observations, 
this is likely an incorrect association. About a decade later, the same region was observed with the~\vla{} four times in 1984 and 1986. \target{} was not detected in the observation made in 1984, likely related to dynamic range limitations. Then, for the observations made in February and July of 1986, \cite{White1989} reported that \target{} was marginally detected (0.69 and 0.71 mJy at C and L band, respectively) only in the first observation. For the observations made in August 1986, \cite{Jackson1987} established that \target{} is a radio-loud source at 1.4 and 4.8 GHz. They report Stokes~I flux densities of $0.69 \pm 0.13$ and $0.51 \pm 0.13$~mJy, respectively. In all observations made in 1986, no significant circularly polarized emission (Stokes~V) or time-variability of the radio emission was detected.
More recently, two observations using J\vla{} and~\lofar{} were made in 2015 and 2017 at 1.6 and 0.14 GHz, respectively. Although the J\vla{} observed a flux density emission of $0.38 \pm 0.02$~mJy from~\target{}, there was no significant detection of circularly polarized emission. \target{} was not detected in the LOFAR observation~\citep[part of the LOFAR Two-metre Sky Survey, LoTSS;][]{Shimwell2017} at 144 MHz, which could be caused by local dynamic range limitations in the presence of the bright radio source (\neighbor{}) that was used as the reference source in this study (Figure~\ref{fig_vla_lofar}). Additional details of the LOFAR observation are presented in Appendix~\ref{app:lofar}.
Finally, preliminary results from the The Karl G. Jansky Very Large Array Sky Survey~\citep[VLASS\footnote{\url{https://science.nrao.edu/science/surveys/vlass}};][]{Lacy2019} show that there was no detection of either dwarfs in~\binary{} at S band for the observations made in 2017, likely associated with low effective integration time (5 seconds).

\subsection{Probability of observing during a non-flaring event} \label{app_probability}

 Given our time sampling, it might be thought that we only detected~\target{} by chance, even though~\companion{} was radio active too. To assess the possibility that in 11 radio observations made at different times (Table~\ref{tab:observations}) we observed~\companion{} during non-flaring states, we assume that flare events follow a Poisson distribution~\citep{Gehrels1986,Pettersen1989} and~\companion{} has similar flaring properties (e.g., rates, flux density) to those of~\target{}. Under this assumption, we find that the probability of detecting~\companion{} at least once in those 11 observations should be $>99\%$. Therefore, since we do not detect~\companion{}, it is unlikely that~\companion{} is as radio active as~\target{}.

\subsection{Comparison of stellar properties with respect to a similar binary system}

The stellar system formed by BL Ceti (also known as Luyten 726-8 A or GJ 65A) and UV Ceti (also known as Luyten 726-8 B or GJ 65B) is one of the most studied dwarf binary systems with flare emissions, where the two stars are similar, and one of the stars is radio loud while the companion remains radio undetectable for some observations. The physical processes behind the flare emission are found to be different for each star, as flares in BL Ceti seem to be radio loud and highly circularly polarized, implying coherent emission mechanisms, whereas UV Cet shows polarized gyrosynchrotron flares and periodic coherent bursts~\citep{Benz1998,Bingham2001,Villadsen2019,Zic2019}. However, if we compare BL-UV Ceti with~\binary{}, the latter presents several particular characteristics, which 
we highlight below.

First, the substantial X-ray emission difference between BL Ceti and UV Ceti has been taken as one of the factors that could be related to the different radio emission due to chromospheric activity~\citep[see, e.g.,][]{Audard2003}. Given this, we have recalculated the X-ray emission for~\target{} and~\companion{} from XMM Newton and Chandra observations (see Table~\ref{t_sourcesinfo}). It has been found that the dwarfs in~\binary{} are equally bright in X-rays, and therefore it is unlikely that there is a one-to-one relation between X-ray and radio activity for the two stellar components.
 
Second, since the binary system BL-UV Ceti is closer to the Sun than~\binary{}, it offers higher angular resolution observations. However, although their orbital separation of $\sim$5AU is too large for a shared magnetosphere, stellar wind interactions are possible, and hence there has been a tentative claim of sympathetic flaring~\citep{Panagi1995b}.
In contrast, since~\binary{} has a separation that is 100 times greater than that of the binary system BL-UV Ceti, any energy flux between the two stars due to stellar wind or other magnetic processes is reduced by a factor of $10^4$, making it highly unlikely that binary magnetic interaction causes detectable effects in the stellar activity.
 
Third, \cite{Berger2002} has studied the flaring radio activity in rotating M and L stars, concluding that the rotation is crucial in determining the physical process behind the flare emission. In this sense, \cite{Audard2003} first suggested that for BL-UV the stellar rotation could play an important role, as BL Ceti (referred as UV Ceti A) is a rapid rotator ($V \ sin \ i = 58$ $\rm{km \ s^{-1}}$), while the companion lacked rotation measurements at that time. However, \cite{Barnes2017} have recently measured rotation periods for the BL-UV Ceti binary system finding that both stars are rapid rotators with similar values (i.e., $V \ sin \ i = 28.6 \pm 0.2$ and $\rm{32.2 \pm 0.2 \ km \ s^{-1}}$). For~\binary{}, discrepancies in the rotation period measurements have also been reported. For~\companion{}, \cite{Kiraga2007} reported a rotation period of 0.950 days ($V \ sin \ i \sim 28.5$ $\rm{km \ s^{-1}}$) based on photometric observations, which was debated by~\cite{Houdebine2015} who obtained $\rm{1.86 \ km \ s^{-1}}$ using a cross-correlation technique for a selected narrow spectral range. This same technique was also used by~\cite{Lopez-Santiago2010} and~\cite{Houdebine2017} obtaining 1.0 and $\rm{6.30 \ km \ s^{-1}}$, respectively. Moreover, using high-resolution spectra, \cite{Reiners2012} reported $\rm{<4.0 \ km \ s^{-1}}$, which was later measured by~\cite{Moutou2017}, who reported $\rm{3.20 \ km \ s^{-1}}$. 
 
For~\target{}, an upper limit of $\rm{\le 10 \ km \ s^{-1}}$ was established by~\cite{Jenkins2009}, which was later refined by~\cite{Jeffers2018} to $\rm{6.79 \ km \ s^{-1}}$. If we do not consider the rotation velocity measured by~\cite{Kiraga2007}, \target{} and~\companion{} can both be categorized as slow rotators (different to what has been found for BL Ceti); however, we cannot confirm this similarity between the dwarfs given the different values reported, and the unknown inclination angles ($i$) that affect the rotation velocity estimates. Additional high-resolution spectra of~\binary{} are necessary to confirm this hypothesis.

Fourth, another speculation suggested for the changes observed in radio activity in M-dwarf binary systems~\citep[such as NLTT33370 and BL-UV Ceti,][ respectively]{Williams2015,Kochukhov2017} is related to the differences in magnetic morphology. For late M dwarfs ($\ge$M6), there is evidence that they can have two different types of magnetic field morphology, i.e., a strong global dipole field or a weaker multipolar field~\citep{Morin2010,Gastine2013}. The strong dipole field might play a role in generating radio emission \linebreak ~\citep[for modeling and observations, see, e.g.,][]{Nichols2012,Hallinan2015,Kao2015,Leto2016,Kuzmychov2017,Turnpenney2017}, but this dynamo bistability so far has only been observed for late M dwarfs. 
 Given that the stars in~\binary{} are both mid-M dwarfs, they could provide a test of the theory that magnetic morphology indeed causes the differences in the radio emission observed. In this sense, the binary system~\binary{} is a potential target for spectropolarimetric observations (Zeeman Doppler Imaging) in order to determine the role of the magnetic morphology in the radio activity observed in the stellar atmosphere of mid-M dwarfs. Moreover, considering the differences in stellar mass between the dwarfs in~\binary{}, it is possible that~\target{} has a fully convective core, while the core of~\companion{} is partly radiative and partly convective (see Section~\ref{nature}). This situation will drive a different dynamo that could lead to different magnetic field topologies~\citep{Browning2007,Yadav2016}, and therefore, different radio activity~\citep{Mclean2012}.


\section{Summary and conclusions}
\label{sec:conclusions}

The radio emission from~\target{} was serendipitously rediscovered by us while inspecting~\gmrt{} data at~$\sim$\amin{8} from the field center~\citep[galaxy cluster~\cluster{},][]{Giacintucci2014,Savini2019}. Comparing two separate observations at 610~MHz from 2007 and 2011, we found that one of the radio sources shifted its position between the epochs by about~\asec{2}, and its flux density changed by a factor of $\sim$4. After ruling out any astrometric or flux density calibration errors, we found that the variable radio emission coincided with the sky position of~\target{} at the times of observation.

We noted that~\target{} forms part of a binary system with \companion{}. The two stars share the same stellar origin and are identical in terms of spectral type, age, and high 3D motion, and infrared, optical, and X-ray emission. However, after inspecting radio observations of this system, ranging from 1984 to 2017, \companion{} remains undetected; it was confirmed to be improbable that~\companion{} was not flaring at any observation. In contrast, \target{} displays radio variability likely associated with a coherent process given the radio circular polarized emission detected, and the brightness temperature estimated ($\rm{>0.7\times10^{12} \ K}$). 

Dwarf binaries with similar companions showing radio flares (where one of the stars is radio loud, while the companion remains radio undetectable) are limited to two known cases: NLTT33370 and BL-UV Ceti~\citep[respectively]{Williams2015,Kochukhov2017}. Nevertheless, the binary system~\binary{} presents a case with closer stellar features between the dwarfs when it is compared to BL-UV Ceti and NLTT33370; the most remarkable difference is their rotational component, which lacks convincing measurements.
Moreover, given its large orbital separation ($>180$ AU), we can discard any magnetic or tidal interaction that could contaminate the independent physical interpretation of the radio activity present in each dwarf. We conclude that there are two intriguing possibilities for the difference in radio activity like that observed in~\binary{}, namely a different magnetic field topology or a vastly different dynamo. They are linked to the rotation component that is measured as $V \ sin \ i$, so the Rossby number would be a more appropriate parameter given the lack of measurements for the inclination angle ($i$)~\citep[see, e.g.,][]{Mclean2012}. Finally, by observing~\binary{}, there is a unique opportunity to disentangle the stellar properties that are linked to the flare emission at radio wavelengths. Further wide-band and higher angular resolution observations of the binary system~\binary{} are already scheduled, and their results are expected to be promising for the radio stellar field.


\begin{acknowledgements}
We sincerely thank the anonymous referee for making valuable suggestions that have improved the paper.
In addition, we thank the staff of the \gmrt{} that made these observations possible. \gmrt{} is run by the National Centre for Radio Astrophysics of the Tata Institute of Fundamental Research.
LOFAR is the Low Frequency Array designed and constructed by ASTRON. It has observing, data processing, and data storage facilities in several countries, which are owned by various parties (each with their own funding sources), and which are collectively operated by the ILT foundation under a joint scientific policy. The ILT resources have benefitted from the following recent major funding sources: CNRS-INSU, Observatoire de Paris and Universit\'{e} d'Orl\'{e}ans, France; BMBF, MIWF-NRW, MPG, Germany; Science Foundation Ireland (SFI), Department of Business, Enterprise and Innovation (DBEI), Ireland; NWO, The Netherlands; The Science and Technology Facilities Council, UK; Ministry of Science and Higher Education, Poland. This research made use of the Dutch national e-infrastructure with support of the SURF Cooperative (e-infra 180169) and the LOFAR e-infra group. The J\"{u}lich LOFAR Long Term Archive and the German LOFAR network are both coordinated and operated by the J\"{u}lich Supercomputing Centre (JSC), and computing resources on the Supercomputer JUWELS at JSC were provided by the Gauss Centre for Supercomputing e.V. (grant CHTB00) through the John von Neumann Institute for Computing (NIC). The University of Hertfordshire high-performance computing facility and the LOFAR-UK computing facility located at the University of Hertfordshire and supported by STFC [ST/P000096/1]. This work also has made use of data from the European Space Agency (ESA) mission \gaia{}\footnote{\url{https://www.cosmos.esa.int/gaia}}, processed by the \gaia{} Data Processing and Analysis Consortium (DPAC\footnote{\url{https://www.cosmos.esa.int/web/gaia/dpac/consortium}}). Funding for the DPAC has been provided by national institutions, in particular the institutions participating in the~\gaia{} Multilateral Agreement. This research has made use of the SIMBAD database,
operated at CDS, Strasbourg, France. L.H.Q.-N. deeply thanks Dr. L.O.~Sjouwerman at NRAO Socorro for the support reducing archival~\vla{} observations and Dr. Ken Croswell for his comments and suggestions.
\end{acknowledgements}


\bibliographystyle{aa}
\bibliography{ross867}


\begin{appendix}

\section{SPAM pipeline for GMRT data on Ross 867-868}
\label{SPAM}
The SPAM pipeline converts the observations into Stokes~\stokesi{} visibilities in an early stage of the processing. This conversion is done on a best effort basis, meaning that if \stokesrr{} and \stokesll{} are both available, Stokes~\stokesi{} is formed through \stokesi{}=(\stokesrr{}+\stokesll{})/2, while if just one of them is available, \stokesi{}=\stokesrr{} or \stokesi{}=\stokesll{}. Using the pipeline, we generated calibrated Stokes~\stokesi{} visibilities and images for all observations.

In addition, the SPAM pipeline derives flux density, bandpass, and instrumental phase calibrations from a primary calibrator (typically 3C\,48 or 3C\,147) and applies them to the target field data (in our case \cluster{}). Then it self-calibrates and images the target field data several times, initially bootstrapping to an externally supplied radio sky model, and ultimately applies SPAM ionospheric calibration to correct for direction-dependent phase errors~\citep[for details, see][]{Intema2009}. For the initial (phase-only) self-calibration of the 235~MHz observations, we bootstrapped to a radio sky model derived from the \gmrt{} 150~MHz all-sky survey~\citep[TGSS;][]{Intema2017}. For every higher frequency, we bootstrapped to an image model derived from the frequency just below it (e.g., a source model from the 235~MHz image was used for bootstrapping the 325~MHz observations). 
This has proven to work well because of the relatively small distance in frequency between adjacent bands and the simple nature of the majority of radio sources, generally providing a good match in terms of resolution, flux density, and sensitivity. Also, the larger field of view at the lower frequency guarantees a model that fully encloses the higher frequency observation.

\section{LOFAR observations}\label{app:lofar}

The LOFAR data was observed as part of the LOFAR Two-metre Sky survey~\citep[LoTSS;][]{Shimwell2016,Shimwell2018}, pointing P260+28 in project ID LC7$\_$024. Initially the data were processed on LOFAR archive compute facilities~\citep[e.g.,][]{Mechev2017} using PreFactor which is the standard LOFAR direction independent calibration pipeline~\citep[see][]{VanWeeren2016,Williams2016}, which corrects for direction independent effects such as the bandpass~\citep[e.g.,][]{DeGasperin2018}. After this, direction dependent calibration was performed to remove the ionospheric errors that are severe at low radio frequencies and to correct for errors in the LOFAR beam model. The direction dependent calibration was performed using the LoTSS processing pipeline\footnote{\url{https://github.com/mhardcastle/ddf-pipeline}} that makes use of kMS~\citep[][]{Tasse2014,Smirnov2015} and DDFacet~\citep[e.g.,][]{Tasse2017} for calibration and imaging while applying direction dependent solutions.
%

\section{Radio survey observations for~\neighbor{}}
\begin{table*}[ht!]
\caption{Observation details for the source~\neighbor{} (\amin{0.8} west of~\target{}) reported in the NVSS and FIRST surveys. The source~\neighbor{} was used as a reference source.\label{t_first_nvss}}
\begin{center}
\begin{tabular}{lcccccc}
\hline \hline
Survey & $\alpha, \delta$\tmark[(3)] & Flux density & Major\tmark[(4)] & Minor\tmark[(4)] & PA\tmark[(4)] & Mean Epoch\tmark[(5)] \\
 & (J2000) & (mJy) & (arcsec) & (arcsec) & (degrees) & (year) \\
\hline
FIRST\tmark[(1)] & \thms{17}{19}{49.314} \adms{+26}{30}{07.69} & $17.9 \pm 0.9$ & 5.52 & 5.37 & 172.8 & 1995.874 \\
NVSS\tmark[(2)] & \thms{17}{19}{49.24} \adms{+26}{30}{07.2} & $17.3 \pm 0.6$ & <19.0 & <18.3 & - & 1995$\pm$2.0 \\
\hline
\end{tabular}
\end{center}
{\footnotesize {\bf Notes.} (1) The Faint Images of the Radio Sky at Twenty-Centimeters (FIRST) survey was made using the NRAO Very Large Array (\vla{}) producing images with 1.8" pixels, a typical rms of 0.15 mJy, and a resolution of 5 arcseconds~\citep{White1997}. (2) The NRAO VLA Sky Survey (NVSS) was made at 1.4 GHz with a resolution of 45 arcseconds and a limiting peak source brightness of about 2.5 mJy/beam~\citep{Condon1998}. (3) Source position right ascension ($\alpha$) and declination ($\delta$) in sexagesimal notation. (4) Source major axis, minor axis, and position angle measured directly from map before deconvolving synthesized beam. No PA reported for NVSS observation.}
\end{table*}

\section{Radio observations for \binary{}}
\begin{table*}[h!]
\caption{Historical observation details for the field of view that contains~\binary{}.\label{tab:observations}}
\begin{center}
\resizebox{\hsize}{!}{ 
\begin{tabular}{ccccccccl}
\hline \hline
 Date Observation & Telescope & Frequency & Bandwidth & Correlations & Time & Sensitivity\tmark[(1)] &$\rm{S_{\nu}}$\tmark[(2)] & Resolution\tmark[(3)] \\
(YYYY-MM-DD) & & (MHz) & (MHz) & parameters & (min) & $\rm{( \mu Jy})$ & (mJy) &\\
\hline
1984-06-29 & \vla & 4,860 & 100 & \stokescfull & 141 & 81 & ND\tmark[(4)] & \sbeam{11.7}{4.7} (\adeg{-89}) \\
1986-02-10 & \vla & 1,540 & 50 & \stokescfull & 50 & 250 & NA\tmark[(5)] & NA\tmark[(5)]\\ 
1986-07-10 & \vla & 1,465 & 50 & \stokescfull & 27 & NA\tmark[(5)] & NA\tmark[(5)] & NA\tmark[(5)] \\
1986-08-06 & \vla & 1,464 & 50 & \stokescfull & 24 & NA\tmark[(5)] & NA\tmark[(5)] & NA\tmark[(5)] \\
 & & 4,760 & 50 & \stokescfull & 15 & NA\tmark[(5)] & NA\tmark[(5)] & NA\tmark[(5)] \\
1990-08-27 & \vla & 1,565 & 100 & \stokescfull & 67 & 320 & ND\tmark[(4)] & \sbeam{4.2}{4.0} (\adeg{-65}) \\
 & & 4,785 & 100 & \stokescfull & 64 & 38 & $0.20\pm0.04$ & \sbeam{1.32}{1.20} (\adeg{-45}) \\
 & & 8,515 & 100 & \stokescfull & 64 & 42 & ND\tmark[(4)] & \sbeam{0.69}{0.68} (\adeg{-27}) \\
2007-03-08 & \gmrt & 325 & 32 & \stokeschalf & 280 & 356 & $5.3\pm0.8$ & \sbeam{9.7}{8.4} (\adeg{-53}) \\
 & & & & \stokesll & & 224 & $11.6\pm1.2$ & \sbeam{10.3}{8.4} (\adeg{-36}) \\
 & & & & \stokesrr & & 248 & $3.5\pm0.6$ & \sbeam{8.9}{8.6} (\adeg{-23}) \\
 2007-03-10 & \gmrt{} & 610 & 32 & \stokeschalf{} & 310 & 52 & $1.7\pm0.2$ & \sbeam{4.5}{3.9} (\adeg{37}) \\
& & & & \stokesll{} & {} & 82 & $2.4\pm0.3$ & \sbeam{4.6}{4.0} (\adeg{41}) \\
& & & & \stokesrr{} & {} & 94 & $0.9\pm0.2$ & \sbeam{4.5}{4.2} (\adeg{13}) \\
2011-07-24 & \gmrt{} & 235 & 17 & \stokesll{} & 346 & 350 & $6.8\pm0.6$ & \sbeam{12.4}{10.7} (\adeg{59}) \\
& & 610 & 33 & \stokesrr{} & & 40 & $7.8\pm0.8$ & \sbeam{5.0}{4.1} (\adeg{77}) \\
2015-03-22 & \jvla{} & 1,519 & 64 & \stokescfull{} & 40 & 15 & $0.38\pm0.02$ & \sbeam{1.32}{1.20} (\adeg{-42})\\
2017-01-25 & \lofar{} & 145 & 48 & \stokeslfull{} & 8 & 200 & ND\tmark[(4)] & \sbeam{8.78}{6.43} (\adeg{62}) \\
2017-10-02 & \jvla{}{} & 2,988 & 150 & \stokescfull{} & 0.83 & 124 & ND\tmark[(4)] & \sbeam{2.56}{2.14} (\adeg{52})\\
\hline
\end{tabular}}
\end{center}
{\footnotesize {\bf Notes.} (1) Sensitivity at the image center. (2) Flux density detected for~\target{}. Where all correlation parameters were available, the value reported corresponds to Stokes~\stokesi{}. (3) Restoring beam major~$\times$~minor axis (position angle). (4) No detection. (5) Not available. Some observations made in 1986 with the~\vla{} did not have a flux calibrator, and were not well centered on the target. Hence, any flux measurement is hard to interpret. We relied on what~\cite{Jackson1987} and~\cite{White1989} reported for these observations (see Section~\ref{sec:historical}).}
\end{table*}

\end{appendix}
%

\end{document}